\documentclass{elsart}
\usepackage{epsfig,amssymb}

\begin{document}
\newcommand{\beqn}{\begin{equation}}
\newcommand{\eeqn}{\end{equation}}
\newcommand{\etag}{\eta\gamma}
\newcommand{\ee}{e^+e^-}
\newcommand{\pipi}{\pi^+\pi^-}
\newcommand{\ompi}{\omega\pi^0}
\newcommand{\pig}{\pi^0\gamma}
\newcommand{\pipig}{{\pi^0\pig}}
\newcommand{\trpi}{\pi^+\pi^-\pi^0}
\newcommand{\rhop}{{\rho\prime}}
\newcommand{\eepipig}{\ee \to \pipig}
\newcommand{\etapig}{\eta\pi^0\gamma}
\newcommand{\fzero}{{f_0(600)}}
\newcommand{\br}{{\mathcal B}}
\newcommand{\rhobr}{(5.2^{+1.5}_{-1.3}\pm 0.6)\times 10^{-5}}
\newcommand{\ombr}{(6.4^{+2.4}_{-2.0}\pm 0.8)\times 10^{-5}}
\newcommand{\rhosibr}{(6.0^{+3.3}_{-2.7}\pm 0.9)\times 10^{-5}}
\newcommand{\ombrli}{3.3 \times 10^{-5}}
\begin{frontmatter}
\date{}

\title{\large \bf \boldmath 
Study of the Process $\eepipig$ in c.m. Energy Range
600--970~MeV at CMD-2}

\author[BINP]{R.R.~Akhmetshin},
\author[BINP,NGU]{V.M.~Aulchenko},
\author[BINP]{V.Sh.~Banzarov},
\author[PITT]{A.~Baratt},
\author[BINP,NGU]{L.M.~Barkov},
\author[BINP]{S.E.~Baru},
\author[BINP]{N.S.~Bashtovoy},
\author[BINP,NGU]{A.E.~Bondar},
\author[BINP]{D.V.~Bondarev},
\author[BINP]{A.V.~Bragin},
\author[BINP,NGU]{S.I.~Eidelman},
\author[BINP]{D.A.~Epifanov},
\author[BINP,NGU]{G.V.~Fedotovitch},
\author[BINP]{N.I.~Gabyshev},
\author[BINP]{D.A.~Gorbachev},
\author[BINP]{A.A.~Grebeniuk}, 
\author[BINP]{D.N.~Grigoriev},
%\author[YALE]{V.W.~Hughes}~\footnote{deceased},
\author[BINP]{F.V.~Ignatov},
\author[BINP]{S.V.~Karpov},
\author[BINP]{V.F.~Kazanin},
\author[BINP,NGU]{B.I.~Khazin},
\author[BINP,NGU]{I.A.~Koop},
\author[BINP]{P.P.~Krokovny},
\author[BINP,NGU]{A.S.~Kuzmin},
\author[BINP]{I.B.~Logashenko},
\author[BINP]{P.A.~Lukin},
\author[BINP]{A.P.~Lysenko},
\author[BINP]{K.Yu.~Mikhailov},
\author[BINP,NGU]{A.I.~Milstein},
\author[BINP]{I.N.~Nesterenko},
\author[BINP]{V.S.~Okhapkin},
\author[BINP]{A.V.~Otboev},
%\author[BINP,NGU]{A.V.~Pak},
\author[BINP,NGU]{E.A.~Perevedentsev},
\author[BINP]{A.A.~Polunin},
\author[BINP]{A.S.~Popov},
\author[BINP]{S.I.~Redin},
\author[BINP]{N.I.~Root},
\author[BINP]{A.A.~Ruban},
\author[BINP]{N.M.~Ryskulov},
\author[BINP]{A.G.~Shamov}, 
\author[BINP]{Yu.M.~Shatunov},
\author[BINP,NGU]{B.A.~Shwartz},
\author[BINP,NGU]{A.L.~Sibidanov},
\author[BINP]{V.A.~Sidorov}, 
\author[BINP]{A.N.~Skrinsky},
\author[BINP]{I.G.~Snopkov},
\author[BINP,NGU]{E.P.~Solodov},
\author[BINP]{P.Yu.~Stepanov},
\author[PITT]{J.A.~Thompson}, 
\author[BINP]{A.A.~Valishev},
\author[BINP]{Yu.V.~Yudin},
\author[BINP,NGU]{A.S.~Zaitsev},
\author[BINP]{S.G.~Zverev}

\address[BINP]{Budker Institute of Nuclear Physics, 
  Novosibirsk, 630090, Russia}
\address[NGU]{Novosibirsk State University, 
  Novosibirsk, 630090, Russia}
\address[PITT]{University of Pittsburgh, Pittsburgh, PA 15260, USA}
%\address[YALE]{Yale University, New Haven, CT 06511, USA}
%========================================================================

\begin{abstract}
The cross section of the process $\eepipig$ has been 
measured in the c.m. energy range 600--970~MeV with the 
CMD-2 detector. The following branching ratios have been determined:
$\br(\rho^0\to\pipig)=\rhobr$ and
$\br(\omega\to\pipig)=\ombr$.
Evidence for the $\rho^0 \to \fzero\gamma$ decay has been obtained: 
$\br(\rho^0 \to \fzero\gamma)=\rhosibr$.
From a search for the process $\ee \to \etapig$
the following upper limit has been obtained:
$\br(\omega \to \etapig) < \ombrli$ at 90\%~CL.
\end{abstract}
\end{frontmatter}
\maketitle

\section{Introduction}
Radiative transitions of vector mesons into two pseudoscalar
mesons have been attracting attention since long ago as a possible 
test of various low energy theoretical models and a source
of information on controversial scalar
states~\cite{Sin62,Ren69,fo90,bgp92,Pra94,mhot99,belmn01,gs01,gsy01,pho02,gsy03,gky03,ok03}.
After the reliable observation of the $f_0(980)$ and $a_0(980)$ states
in the $\phi(1020)$ meson decays by SND~\cite{sndphi} and 
CMD-2~\cite{cmd2phi}, recently confirmed by KLOE~\cite{kloe}, 
the interest moved to the $\rho$ and $\omega$ meson decays. 

Information on the $\rho(\omega)$ decays to the 
$\pi\pi\gamma(\etapig)$ final states is rather scarce. 
Because of the large background from the initial state radiation
in the process $e^+e^- \to \pi^+\pi^-$, a search for $\rho(\omega)$
decays into the $\pi^+\pi^-\gamma$ final state is difficult and among
many such experiments~\cite{pdg} only one succeeded in the
observation of the decay $\rho^0 \to  \pi^+\pi^-\gamma$~\cite{ndrho}.
A long search for the $\omega\to\pipig$ decay (see~\cite{pdg}
and references therein) finally proved successful for the
GAMS Collaboration, which observed it in $\pi^-p$ collisions with the 
branching fraction of $(7.4 \pm 2.5) \times 10^{-5}$~\cite{gams}. 
Recently a high statistics study of the $\rho(\omega)$ energy range 
has been performed by the SND Collaboration~\cite{snd1,snd2}.
They observe both  $\rho$ and $\omega$ decays 
into $\pipig$ with a branching ratio higher than that predicted
by vector dominance. While for the $\rho$ meson the excess can be
explained by the $\rho^0 \to\fzero\gamma$ decay, first evidence for which is 
reported by SND, the reasons for the higher probability of the 
corresponding $\omega$ decay are not yet clear.         

In our recent paper~\cite{ompi_cmd} we described a study of the
process $\eepipig$ in the c.m. energy range 920--1380~MeV, i.e.
above the threshold of $\ompi$ production, using 
the CMD-2 detector at the VEPP-2M $\ee$ collider.  
In this work we report on the measurement of the cross section 
of the process $\eepipig$ in the c.m. energy range 600--970~MeV
with CMD-2. Also described is 
the first search for the process $\ee\to\etapig$ in this energy range. 
%The preliminary results of this work were published in~\cite{mast}.

\section{Experiment}

The general purpose detector CMD-2 has been described in 
detail elsewhere~\cite{cmddet}. Its tracking system consists of a 
cylindrical drift chamber (DC) and double-layer multiwire proportional 
Z-chamber, both also used for the trigger, and both inside a thin 
(0.38~X$_0$) superconducting solenoid with a field of 1~T. 
The barrel CsI calorimeter with a thickness of 8.1~X$_0$ placed
outside  the solenoid has energy resolution for photons of about
9\% in the energy range from 100 to 700~MeV. The angular resolution is 
of the order of 0.02 radians. The end-cap BGO calorimeter with a 
thickness of 13.4~X$_0$ placed inside the solenoid 
has energy and angular resolution varying from 9\% to 4\% and from 
0.03 to 0.02 radians, respectively, for the photon energy in the range 
100 to 700~MeV. The barrel and end-cap calorimeter systems cover a 
solid angle of $0.92\times4\pi$ radians. 

This analysis is based on a data sample corresponding to
integrated luminosity of 7.7~pb$^{-1}$ collected in 1998--2000 
in the energy range 600--970~MeV.  
The step of the c.m. energy scan varied from 0.5~MeV near the $\omega$
peak to 5~MeV far from the resonance.
The beam energy spread is about $4\times 10^{-4}$ of the total energy.
The luminosity is measured using events of Bhabha scattering 
at large angles~\cite{prep}.

We use Monte Carlo simulation (MC) to model the response of
the detector and determine the efficiency.
Due to the beam background additional (``fake'') clasters can
appear in the calorimeter. 
The corresponding probability as well as photon energy and angular 
spectra are 
obtained directly from the data using the process $\ee\to\pipi\pi^0$. 
Then these photons are
mixed with the detector response during simulation.

\section{Data analysis}

At the initial stage, we select events
which have no tracks in the DC, five photons, the total energy 
deposition $1.7 < E_{\rm tot}/ E_{\rm beam} < 2.2$, the total momentum 
$P_{\rm tot}/ E_{\rm beam}<0.3$ and at least three photons detected in the
CsI calorimeter. The minimum photon energy is 30~MeV for the CsI and 
40~MeV for the BGO calorimeter. 
Then a kinematic fit requiring energy-momentum conservation is
performed with an additional reconstruction of two $\pi^0$ mesons.
We require good reconstruction quality ($\chi^2 < 5$) and
the ratio of the reconstructed to measured energy to be 
$0.75 < \omega_i\, / E_i < 1.8$ for each photon. After this stage
219 events remain in the whole energy range.

The dominant background comes from the processes
\begin{equation}
  \label{bg_etag}
  \ee\to\eta\gamma,~\eta\to 3\pi^0,
\end{equation}
\begin{equation}
  \label{bg_pi0g}
  \ee\to\pi^0\gamma,
\end{equation}
\begin{equation}
  \label{bg_qed}
  \ee\to 3\gamma,~4\gamma.
\end{equation}
Events from the process (\ref{bg_etag}) can imitate signal events
if two soft photons are lost. The processes (\ref{bg_pi0g},\ref{bg_qed})
can meet the selection criteria with one ore two additional (``fake'') 
photons coming from shower splitting, ``noisy'' electronic channels in 
the calorimeter and beam background.

To determine the background contribution,
the following procedure is used.
All above listed processes are simulated using the Monte Carlo. 
To decrease a statistical uncertainty, ten times more events than is
expected from the background cross section and luminosity are used
at each energy. 
Then the same selection criteria as for the data are applied. The obtained
number of selected events is divided by ten and subtracted from
experimental data at each energy. In total, about 29 background events
are expected after this procedure.

\begin{figure}
  \includegraphics[width=0.5\textwidth]{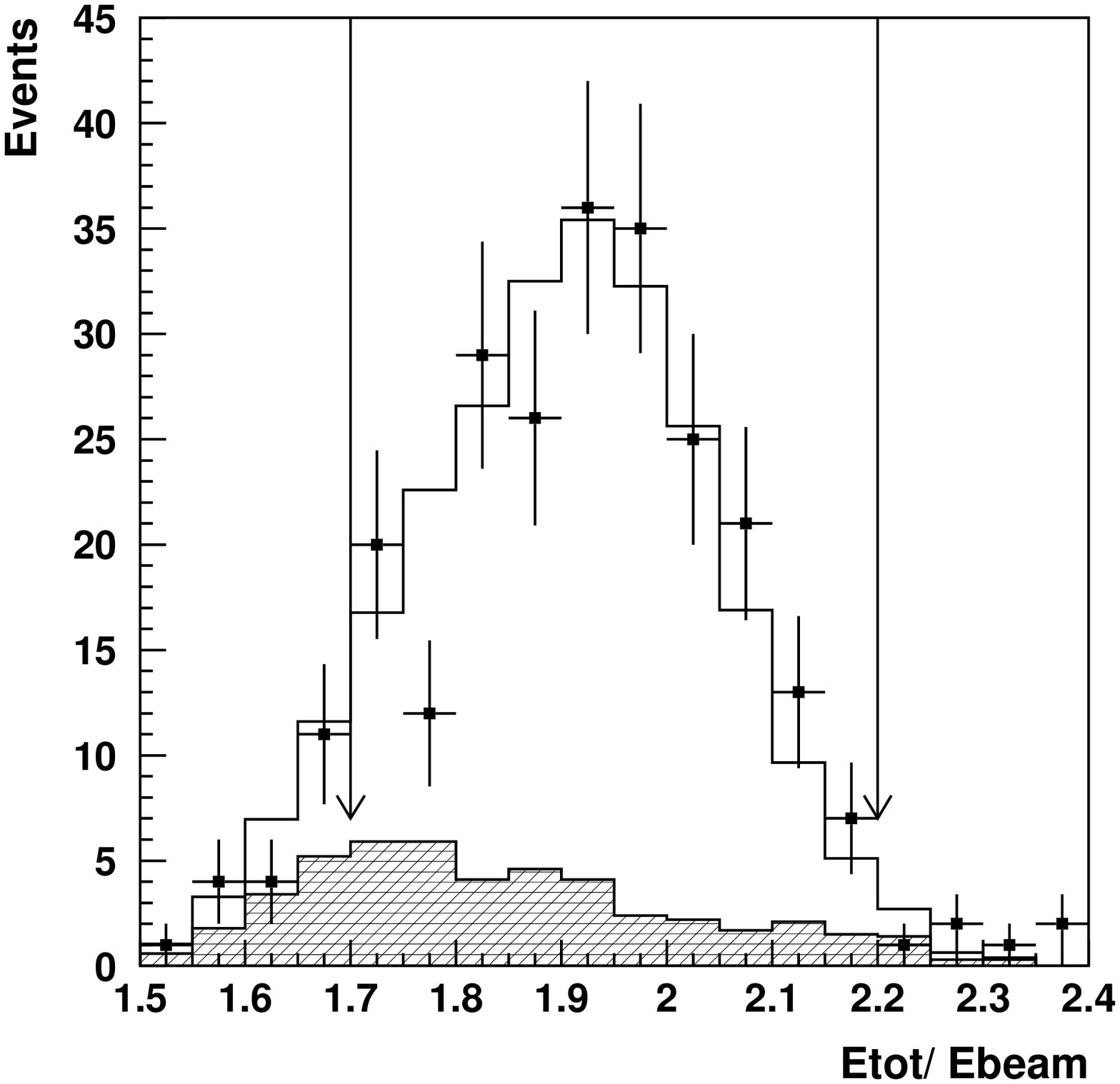}
  \includegraphics[width=0.5\textwidth]{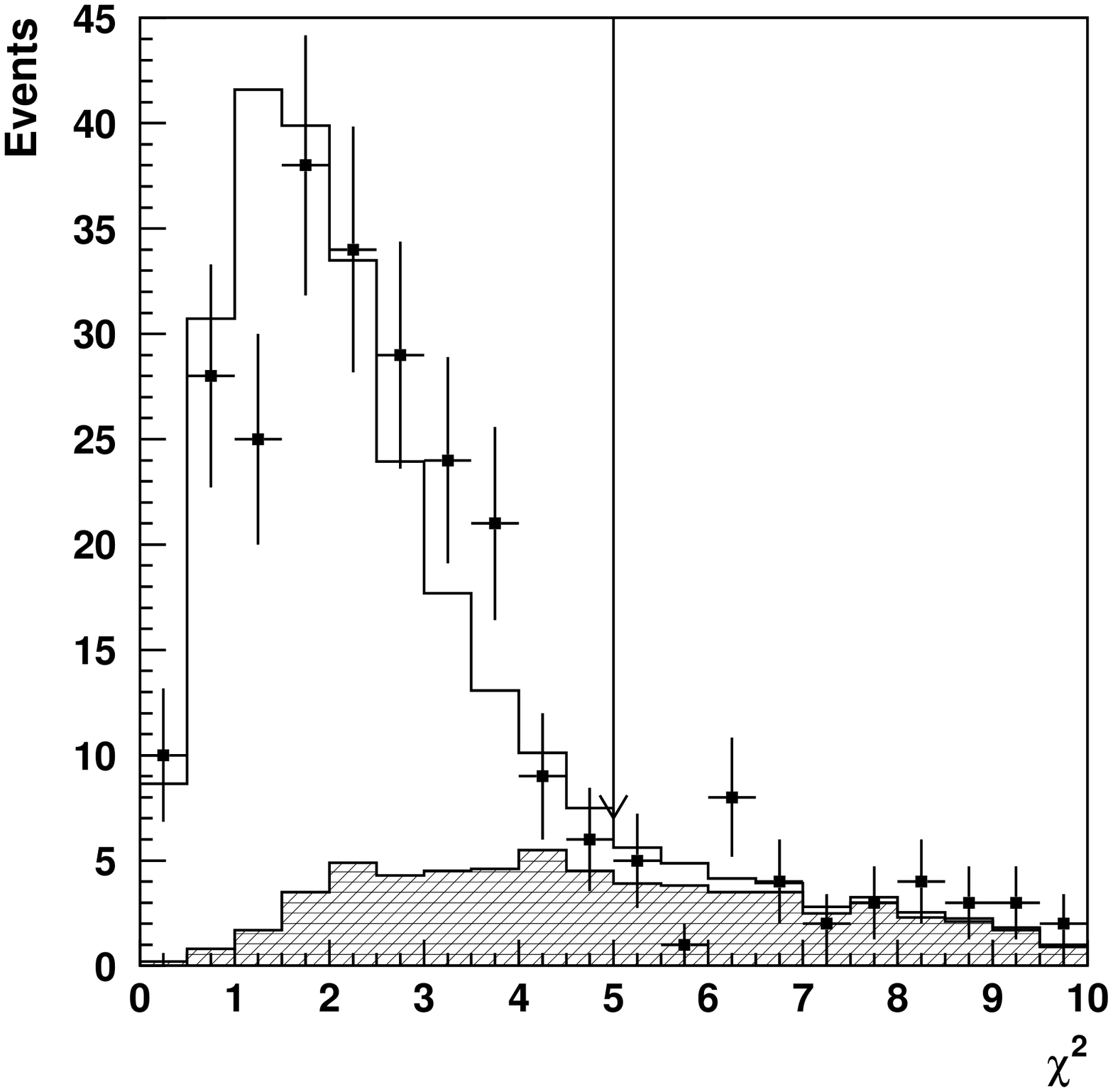}
  \caption{The $E_{\rm tot}$ (left) and $\chi^2$ distributions. The 
    points with error bars represent experimental events, hatched 
    histograms show the MC simulation for the background processes and 
    open histograms are a sum of the signal and background MC.  
    The arrows indicate the cuts imposed.}
  \label{om_comp}
\end{figure}
 
Figure~\ref{om_comp} shows the $E_{\rm tot}$ and $\chi^2$ distribution for 
the data, signal and background MC. We use the $\chi^2$ distributions to
estimate the accuracy of the background estimation. The experimental
distribution was fitted by a sum of MC and background. The ratio
$N_{\rm bg}^{\rm exp}/N_{\rm bg}^{\rm MC}=1.2\pm 0.2$ was obtained. 
We conclude that the background level is estimated well and its 
systematic error does not exceed 40\%. This results in a 6\%
systematic uncertainty in the cross section.

\subsection{Approximation of the cross sections}

At each energy point the cross section of the process $\sigma$ is 
calculated from the observed number of events and background MC
expectation using the following formula:
\begin{equation}
\sigma(\sqrt{s})=\frac{N_{\rm exp}-N_{\rm bg}}{L \varepsilon (1+\delta)}
\, , \label{Nth}
\end{equation}
where $N_{\rm exp}$ is the number of observed events, $N_{\rm bg}$ is the
expected number of background events from MC,
$L$ is the integrated luminosity, 
$\varepsilon$ is the detection efficiency and
$(1+\delta)$ is the radiative correction at the corresponding energy.  

To calculate the detection efficiency,
we use Monte Carlo simulation taking into account the neutral 
trigger (NT) efficiency. NT is based on the information from the CsI 
calorimeter and its efficiency depends on the number of clusters and 
total energy deposition. The NT efficiency is estimated using events of
the process $\ee\to\pi^+\pi^-\pi^0$. 
We require the charged trigger signal and three or more clusters in
the CsI calorimeter, and study the NT efficiency as a function of the
energy deposition in CsI. The NT efficiency varies from 85\% at
c.m. energy 600~MeV to about 95\% at 980~MeV.
%~\cite{mast}.

The radiative corrections are calculated according to~\cite{radcor}.
The dependence of the detection efficiency on the energy 
of the emitted photon is determined from simulation.

\begin{table*}
\vspace*{-9mm}
\caption{The c.m. energy, integrated luminosity, detection efficiency, 
  number of observed events, background expectation, radiative 
  correction, Born cross section $\sigma$, vacuum polarization 
  correction and  ``bare'' cross section $\hat{\sigma}$ of the 
  process $\eepipig$.}
\medskip
\begin{tabular*}{\textwidth}{c@{\extracolsep{\fill}}rcrccrcr}
\hline\hline
$\sqrt{s}$, MeV & $L$, nb$^{-1}$& $\varepsilon$, \% & $N_{\rm exp}$ &
$N_{\rm bg}$ & $1+\delta$& $\sigma$, pb & $|1 - \Pi(s)|^2$ &
$\hat{\sigma}$, pb\\\hline
600&  56.1& 10.9& 0& 0.2& 0.892& $<411$&0.993& $<408$\\
630& 115.1& 11.9& 0& 0.1& 0.888& $<192$&0.995& $<191$\\
660& 235.5& 12.8& 0& 0.4& 0.884& $<80$&0.997& $<80$\\
690& 196.2& 13.5& 0& 0.2& 0.880& $<96$&0.999& $<96$\\
720& 419.7& 14.1& 1& 0.4& 0.879& $<76$&0.999& $<76$\\
750& 210.5& 14.7& 3& 0.1& 0.884& $106^{+84}_{-68}$&0.994& $105^{+83}_{-68}$\\
760& 206.5& 14.9& 2& 0.3& 0.885& $62^{+82}_{-38}$&0.991& $62^{+81}_{-38}$\\
764&  39.7& 14.9& 1& 0.0& 0.883& $191^{+334}_{-120}$&0.990&$189^{+331}_{-119}$\\
770& 109.2& 15.0& 1& 0.3& 0.872& $<284$&0.991& $<282$\\
774& 195.3& 15.1& 4& 0.9& 0.852& $123^{+110}_{-66}$&0.994&$122^{+109}_{-66}$\\
778& 199.1& 15.2& 6& 1.3& 0.817& $190^{+132}_{-88}$&0.994&$189^{+131}_{-87}$\\
780& 194.7& 15.2& 11& 1.1& 0.801& $417^{+161}_{-92}$&0.983&$410^{+158}_{-90}$\\
781& 255.7& 15.2& 5& 1.5& 0.798& $112^{+90}_{-41}$&0.971& $109^{+87}_{-40}$\\
782& 631.0& 15.3& 30& 5.1& 0.803& $322^{+62}_{-59}$&0.958& $309^{+60}_{-57}$\\
783& 275.7& 15.3& 15& 2.6& 0.815& $361^{+126}_{-107}$&0.946&$342^{+119}_{-101}$\\
784& 337.2& 15.3& 16& 3.3& 0.835& $295^{+111}_{-85}$&0.937&$276^{+104}_{-80}$\\
785& 198.8& 15.3& 9& 1.6& 0.859& $283^{+145}_{-102}$&0.932&$264^{+135}_{-95}$\\
786& 190.8& 15.3& 10& 1.4& 0.885& $332^{+147}_{-124}$&0.932&$309^{+137}_{-116}$\\
790& 149.4& 15.4& 4& 0.6& 0.966& $153^{+125}_{-75}$&0.939&$144^{+117}_{-70}$\\
794& 178.7& 15.5& 4& 0.4& 1.000& $130^{+100}_{-60}$&0.944& $123^{+94}_{-57}$\\
800& 261.7& 15.6& 2& 0.7& 1.010& $32^{+55}_{-24}$&0.948& $30^{+52}_{-23}$\\
810& 247.3& 15.8& 3& 1.0& 1.006& $51^{+59}_{-40}$&0.950& $48^{+56}_{-38}$\\
820& 295.7& 15.9& 3& 0.7& 1.001& $49^{+49}_{-34}$&0.951& $46^{+46}_{-32}$\\
840& 602.8& 16.3& 13& 0.8& 0.999& $124^{+44}_{-38}$&0.953& $118^{+42}_{-36}$\\
880& 375.4& 17.1& 4& 0.6& 0.984& $54^{+37}_{-26}$&0.958& $51^{+35}_{-25}$\\
920& 458.2& 18.1& 6& 1.1& 0.901& $65^{+44}_{-29}$&0.964& $63^{+42}_{-28}$\\
940& 327.8& 18.7& 12& 0.5& 0.854& $219^{+82}_{-60}$&0.966& $212^{+79}_{-58}$\\
950& 226.1& 19.1& 14& 0.4& 0.855& $369^{+117}_{-100}$&0.968&$357^{+113}_{-97}$\\
958& 250.0& 19.3& 17& 0.3& 0.859& $402^{+116}_{-101}$&0.969&$390^{+112}_{-98}$\\
970& 249.8& 19.8& 23& 0.9& 0.867& $516^{+116}_{-101}$&0.972&$502^{+112}_{-98}$\\
\hline\hline
\end{tabular*}
\label{ompi_cs}
\end{table*}

\begin{figure*}
\centering
  \includegraphics[width=0.5\textwidth]{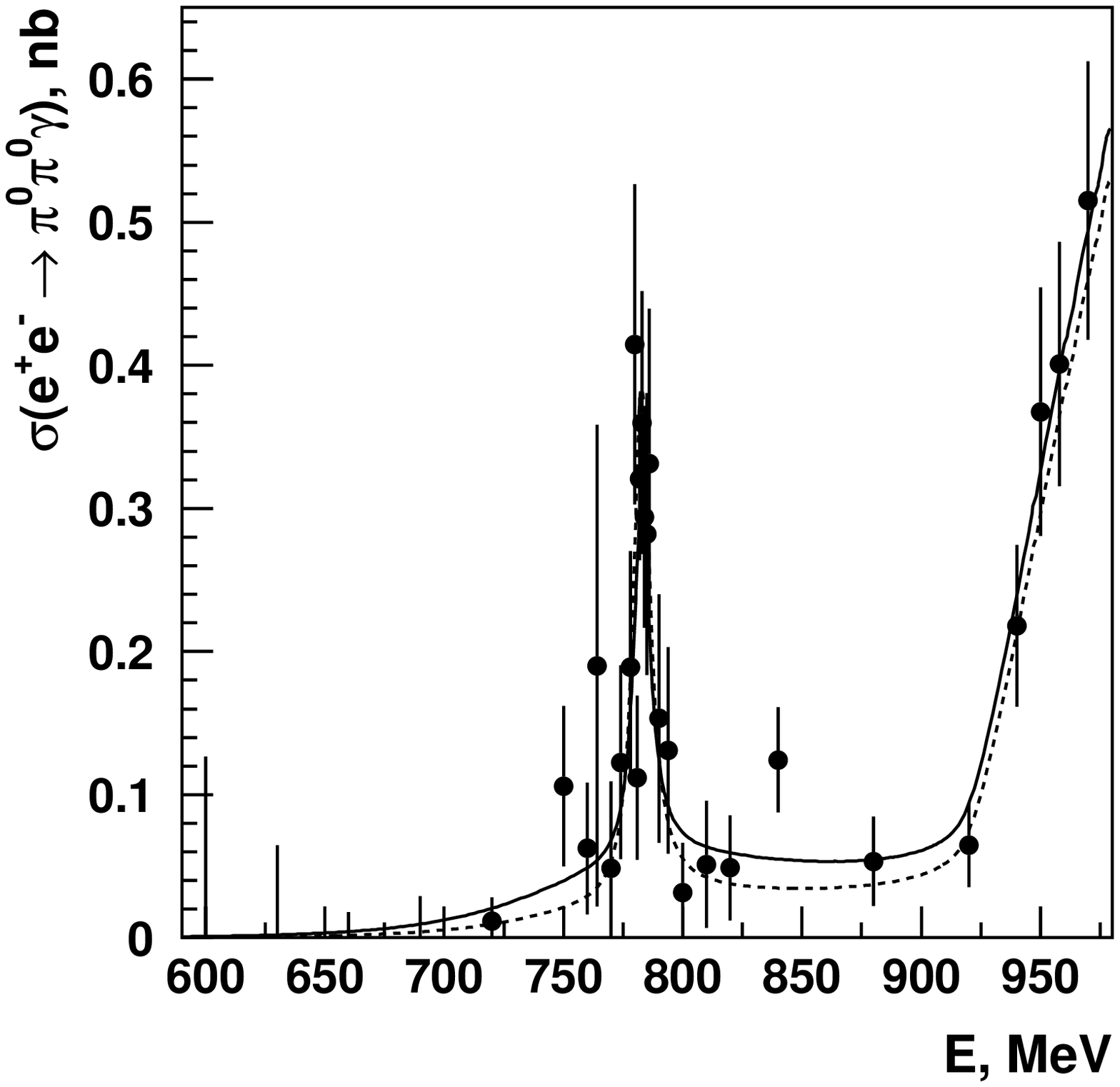}\hfill
  \includegraphics[width=0.5\textwidth]{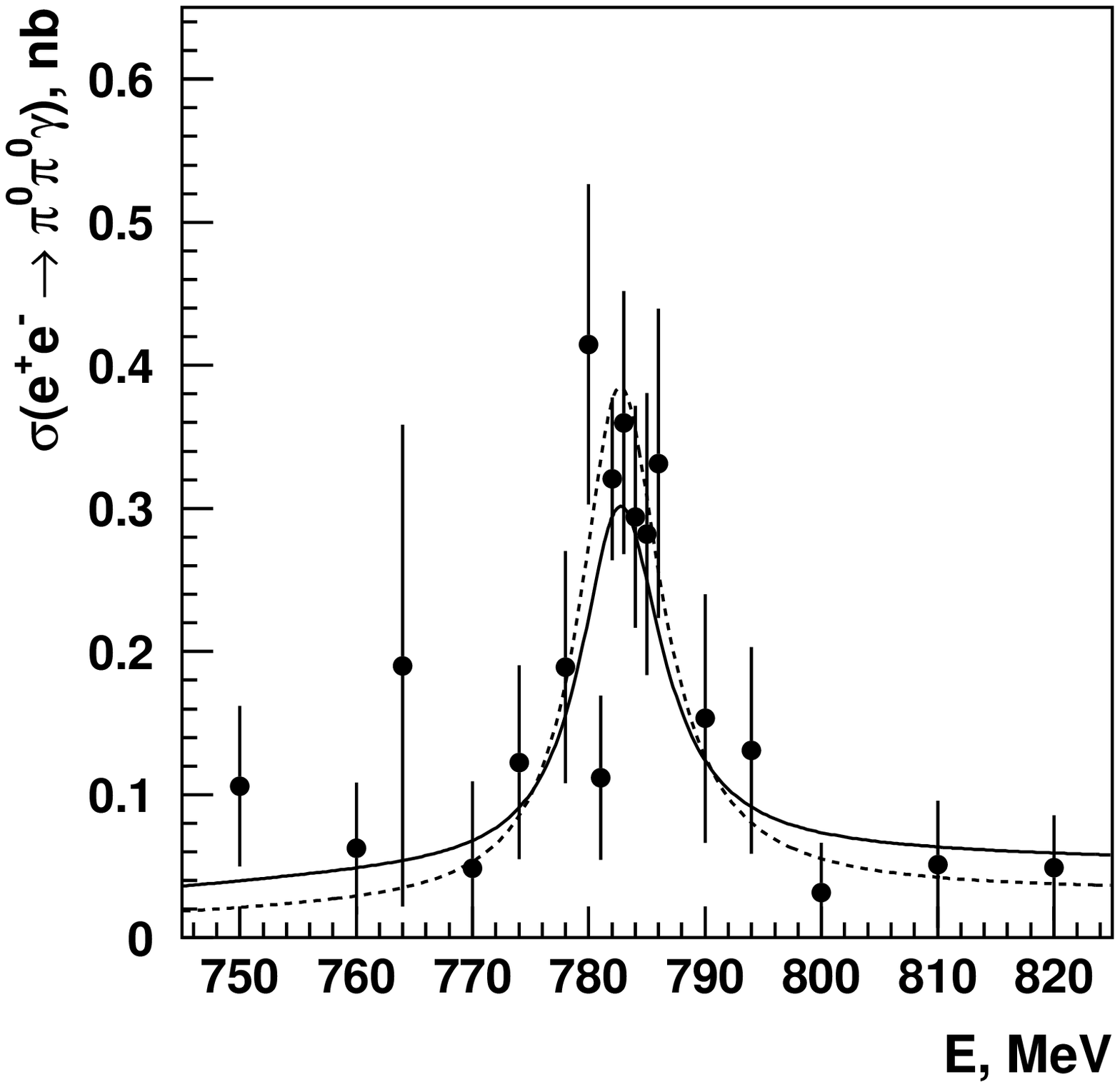}
  \caption{The cross section of the process $\eepipig$. The points
  with error bars represent the experimental data, while the solid and
  dashed curves correspond to the results of the fit I and III, 
  respectively.}
  \label{cs_cmd}
\end{figure*}

The obtained Born cross section of the process 
$\eepipig$ is shown in Fig.~\ref{cs_cmd}
while Table~\ref{ompi_cs} lists detailed information on the analysis
of this reaction.
No events were selected in the energy range 600 to 690~MeV, therefore our
results are presented as upper limits at 90\% C.L.
The Feldman-Cousins procedure~\cite{feldman} was used to calculate
errors (upper limits) at each energy.
This cross section (the ``dressed'' one from the column VII) was
used in the approximation of the energy dependence with resonances. 

Meanwhile, for 
applications to various dispersion integrals like that for the leading
order hadronic contribution to the muon anomalous magnetic moment, 
the ``bare'' cross section should be used. 
Following the procedure in
Ref.~\cite{rho}, the latter is obtained from the ``dressed'' one by
multiplying it by the vacuum polarization correction $|1 - \Pi(s)|^2$,
where  $\Pi(s)$ is the photon polarization operator calculated taking
into account the effects of both leptonic and hadronic vacuum
polarization. The values of the correction and the ``bare'' cross
section $\hat{\sigma}$ are presented in two last columns of 
Table~\ref{ompi_cs}. 

The maximum likelihood method is applied to fit the experimental
data obtained from the relation (\ref{Nth}).
We parameterize the amplitude of the process by a sum of the $\rho$ and
$\omega$ contributions. The former contains the  $\rho\to\ompi$
transition plus one more mechanism beyond the vector dominance model 
which is chosen to be the $\rho^0\to\fzero\gamma$ one. Because of the 
small width of the $\omega$ meson, the $\rho\to\ompi$ amplitude is
rapidly falling below the $\ompi$ threshold, making thereby the
$\rho^0\to\ompi$ and $\rho^0\to\fzero\gamma$ transitions 
distinguishable. On the contrary, the small width of the $\omega$
meson prevents from distinguishing various
mechanisms possibly existing in the  $\omega\to\pipig$ decay. 
For this reason we parameterize the $\omega$ meson amplitude  
by the $\omega\to\rho\pi$ transition only.

The Born cross section of the process is written as:
\begin{equation}
\sigma_{\pipig}(s) = \int{|A_\pipig(s)|^2} d\Phi\; ,
\label{sigma_def}
\end{equation}
where $d\Phi$ is the final state phase space and
\begin{equation}
A_\pipig= A_{\rho\to\ompi}(\frac{m_\rho^2}{D_\rho}+
\alpha\frac{m^2_\rhop}{D_\rhop}) + 
A_{\rho\to\fzero\gamma}\frac{m^2_\rho}{D_\rho} + 
A_{\omega\to\pipig}\frac{m^2_\omega}{D_\omega}\; .
\label{ompi_rho}
\end{equation}
Here the first term describes the amplitude of the $\ee\to\rho$, 
$\rhop\to\ompi$ transition, while the second and third ones are the
$\ee\to\rho\to\fzero\gamma$ and $\ee\to\omega\to\rho\pi^0$ amplitudes.
$m_V$ is the mass and $D_V$ is the propagator of the vector meson $V$ 
given by $D_V(s)=s-m^2_V+i\sqrt{s}\Gamma_V(s)$, $\Gamma_V(s)$ is the
corresponding energy dependent width.
The real parameter $\alpha=g_{\rhop\omega\pi}/g_{\rho\omega\pi}$ is 
the ratio of the coupling constants for the $\rho$ and $\rhop$ mesons.
The $A_{\rho\to\ompi}$ amplitude, proportional to the coupling 
constant $g_{\rho\omega\pi}$ of the $\rho\to\omega\pi$ transition,
is written as in our previous analysis
above 1~GeV~\cite{ompi_cmd}. 

The coupling constant $g_{\rho\omega\pi}$ and the following branching 
ratios are used as fit parameters:
\begin{eqnarray}
  \label{eq:br_def}
\br(\rho^0\to\fzero\gamma\to\pipig)=\frac{1}{\sigma_\rho} \int |
A_{\rho\to\fzero\gamma}(m_\rho)|^2 d\Phi\; ,\\
\nonumber 
\br(\omega\to\pipig)=\frac{1}{\sigma_\omega}\int |
A_{\omega\to\pipig}(m_\omega)|^2 d\Phi\; .
\nonumber
\end{eqnarray} 
Then the total branching fraction of the $\rho^0\to\pipig$ decay 
is calculated from the following formula:
\begin{equation}
  \label{eq:br_rho}
\br(\rho\to\pipig)=\frac{1}{\sigma_\rho} \int |A_{\rho\to\ompi}(m_\rho)+
A_{\rho\to\fzero\gamma}(m_\rho)|^2 d\Phi\; .  
\end{equation}
In (\ref{eq:br_def}) and (\ref{eq:br_rho})
$\sigma_V$ is the cross section at the resonance peak 
without taking into account other contributions:
\begin{equation}
  \sigma_{V}=\sigma_{\ee\to V\to \pipig}(m_V^2)=
  \frac{12\pi \br(V \to \ee) \br(V \to \pipig)} {m^2_V}\; ,
\label{eq:sigma_G}
\end{equation}
$\br(V\to\ee)$ and $\br(V\to\pipig)$ are the corresponding branching 
ratios.

\subsection{Results of the fits} 

In all the following fits the $\rho$, $\omega$ and $\rhop$ meson
masses and widths are fixed at the world average values~\cite{pdg}. 
The $\fzero$ mass and width are badly known~\cite{pdg}. Therefore, we use
a wide range of these parameters:
$M_\fzero=400$--800~MeV, $\Gamma_\fzero=300$--600~MeV.
For the $\rho$ and $\rhop$ resonances the energy dependence
of the total width was described similarly to Ref.~\cite{ompi_cmd}
while for the $\omega$ meson the total width was assumed to be energy 
independent.  
We perform three main fits: with  $g_{\rho\omega\pi}$ equal to the
value $(16.7\pm 0.4\pm 0.6)$~GeV$^{-1}$ obtained in our analysis 
above 1~GeV~\cite{ompi_cmd} (fit I),
with free $g_{\rho\omega\pi}$ (fit II) and without a contribution from
the $\rho^0\to\fzero\gamma$ decay (fit III).
The results of the fits are shown in Table~\ref{BB} and in 
Fig.~\ref{cs_cmd} by the curves.

\begin{table*}
\caption{ The fit results in various models}
\medskip
\begin{tabular*}{\textwidth}{l@{\extracolsep{\fill}}ccc}\hline\hline
Fit parameters      & Fit I  & Fit II & Fit III \\\hline
$\br(\omega\to\pipig)$, $10^{-5}$ & $6.4^{+2.4}_{-2.0}\pm 0.8$ &
$6.2^{+2.4}_{-2.0}\pm 0.7$ & $11.8^{+2.1}_{-1.9}\pm 1.4$\\
$g_{\rho\omega\pi}$, GeV$^{-1}$ & $\equiv 16.7$ & $16.2\pm 1.4$ &
$18.6\pm 1.1$\\
$\br(\rho\to \fzero\gamma\to\pipig)$, $10^{-5}$ &
$2.0^{+1.1}_{-0.9}\pm 0.3$ & $2.3^{+1.4}_{-1.2}\pm 0.3$ & $\equiv 0$\\
$\br(\rho\to\pipig)$, $10^{-5}$ & $5.2^{+1.5}_{-1.3}\pm 0.6$ &
$5.4^{+1.6}_{-1.4}\pm 0.6$ & $2.2\pm 0.3\pm 0.3$\\
\hline
$\chi^2/$ n. d. f. & 19.2 / 28 & 19.0 / 27 & 26.7 / 28\\
\hline\hline
\end{tabular*}
\label{BB}
\end{table*}

The value of the coupling constant
$g_{\rho\omega\pi}=16.2\pm 1.4$  
obtained in the second fit is in good agreement with the one 
from our measurement of the $\ee\to\ompi$ cross section above 
1~GeV~\cite{ompi_cmd}. 

The fits taking into account the $\rho^0\to\fzero\gamma$
decay mode (fits I and II) better describe data. The branching fraction 
$\br(\rho^0\to\fzero\gamma)$ differs from zero by two standard deviations.
However, the fit III also has good $\chi^2/{\rm n.~d.~f.}=26.7/28$. 
There are additional reasons to choose a fit with the 
$\rho\to\fzero\gamma$ decay:
\begin{itemize}
\item The branching fraction of the $\omega\to\pipig$ decay
obtained in the fit III,  $\br(\omega\to\pipig)=
(11.8^{+2.1}_{-1.9}\pm 1.4)\times 10^{-5}$ is above the GAMS result
$(7.4\pm 2.5)\times 10^{-5}$~\cite{gams} by two standard deviations.
The latter result obtained in $\pi^- p$ collisions 
has no $\rho$ contribution.

\item The value of  the coupling constant
$g_{\rho\omega\pi}=18.6\pm 1.1$
  obtained in the fit III is above that from our analysis above 
1~GeV~\cite{ompi_cmd} by almost two standard deviations.

\item The recent analysis of the process $\ee\to\pipig$ by 
SND~\cite{snd2} also showed evidence for
the $\rho^0\to\fzero\gamma$ decay.
\end{itemize}
Thus, we choose the first model as our final result.
The $\rho^0\to\fzero\gamma$ branching fraction is calculated from 
the results listed in Table~\ref{BB} taking into
account that $\br(\fzero\to\pi^0\pi^0)=1/3$.

\subsection{Invariant mass spectrum}

\begin{figure}
\centering
\includegraphics[width=.5\textwidth]{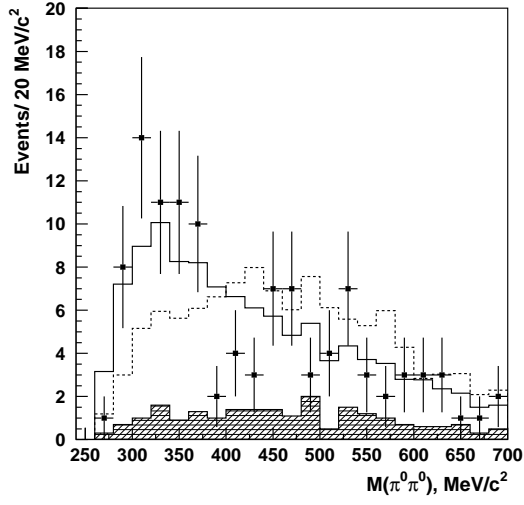}
\caption{The invariant mass of two $\pi^0$ mesons in the $\omega$ meson 
energy range. The points with error bars represent the experimental data,
the solid histogram shows the MC simulation of the $\omega\to\rho\pi^0$ 
decay, the dashed one corresponds to the $\omega\to\fzero\gamma$ 
decay. The hatched histogram is the estimated background
contribution.}
\label{fig:mpipi}
\end{figure}

Figure~\ref{fig:mpipi} shows the spectrum of a $\pi^0\pi^0$ invariant 
mass for experimental events from the $\omega$ meson energy range 
(770--800~MeV). The experimental distribution agrees well with the
$\omega\to\rho\pi^0$ decay model, however, the contribution from the
$\omega\to\fzero\gamma$ decay is also acceptable. The existing
statistics is not enough to distinguish these contributions,
therefore we obtain only the total branching fraction of the 
$\omega\to\pipig$ decay.

\subsection{Search for the decay $\omega\to\etapig$}

\begin{figure}
\centering
  \includegraphics[width=0.5\textwidth]{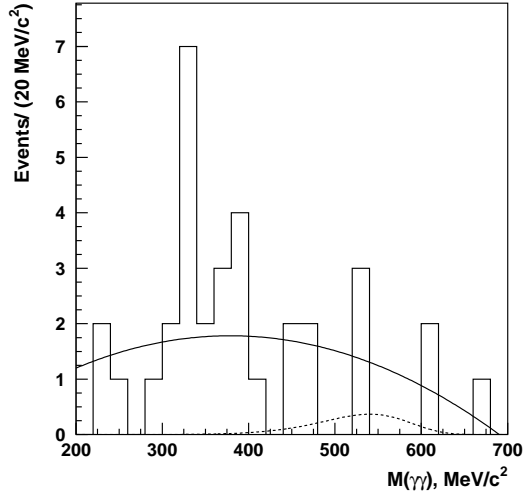}
  \caption{The $M_{\gamma\gamma}$ distribution for the $\etapig$
    candidates. The histogram represents experimental events and
    the solid curve shows the fit result. The dashed curve corresponds
    to the $\etapig$ MC.}
  \label{mgg}
\end{figure}

For a search of events of the process $\ee\to\etapig$ we first apply
the same criteria as for the preliminary selection of $\ee\to\pipig$ 
events. After that a kinematic fit requiring energy-momentum
conservation is performed with the additional reconstruction of 
one soft $\pi^0$ meson and a good reconstruction quality, $\chi^2<6$,
is required. To reject the dominant background from the process 
$\ee\to\pipig$, we perform an additional kinematic fit with the 
$\pipig$ hypothesis and reject events that are consistent with it, 
$\chi^2_{\pipig}<6$. 
Then we look for a possible $\eta$ signal in the invariant mass of 
two hard photons of the remaining three, $M_{\gamma\gamma}$.
The $M_{\gamma\gamma}$ distribution is approximated with a
Gaussian for the signal and a polynomial function for the background.
The Gaussian mean value and width are fixed from the MC simulation of
the signal events. The background shape is obtained using the
$\pipig$ MC. In all energy ranges the resulting $\etapig$ signal
is consistent with zero. Figure~\ref{mgg} shows the
$M_{\gamma\gamma}$ distribution for events from the $\omega$ resonance
region: $381~\mbox {MeV} < E_{\rm beam}< 401~\mbox {MeV}$.
The 90\% CL upper limit for the number of $\etapig$ events is obtained:
$N_{\eta\pig}<2.4$. Using the detection efficiency of 1.3\%,
we set the following upper limit for the $\ee\to\etapig$ cross section: 
\begin{equation}
  \label{etapig_cs}
  \sigma(\ee\to\etapig)<57~\mbox{pb}\, ,
\end{equation}
and for the branching fraction of the $\omega$ meson:
\begin{equation}
  \label{etapig_br}
  \br(\omega\to\etapig)<\ombrli\, .
\end{equation}

\subsection{Systematic errors}

The main sources of  systematic uncertainties in the cross section 
determination are listed in Table~\ref{syst}.
The systematic error due to selection criteria is obtained by varying
the photon energy threshold, total energy deposition, total momentum,
and $\chi^2$.
The model uncertainty corresponds to different detection
efficiencies for the $\ompi$, $\rho\pi^0$ and $\fzero\gamma$
intermediate states. It also includes dependence on the $\fzero$
mass and width.
The uncertainty in the determination of the integrated luminosity 
comes from the selection criteria of Bhabha events, radiative
corrections and calibrations of DC and CsI. The error of the NT 
efficiency was estimated by trying various
fitting functions for the energy dependence and variations of the
cluster threshold. The uncertainty of the radiative corrections comes 
from the dependence on the emitted photon energy and the accuracy of
the theoretical formulae. The resulting systematic uncertainty of the 
cross section in Table~\ref{ompi_cs} as well as of the branching 
fractions in Table~\ref{BB} is 12\%.

\begin{table*}
\caption{Main sources of systematic errors}
\medskip
\begin{tabular*}{\textwidth}{l@{\extracolsep{\fill}}c}
\hline\hline
Source & Contribution, \%\\
\hline
Selection criteria & 8\\
Background subtraction & 6\\
Model uncertainty & 5\\
Luminosity & 2\\
Trigger efficiency & 2 \\
Radiative corrections & 1 \\
\hline
Total & 12\\
\hline\hline
\end{tabular*}
\label{syst}
\end{table*}

\section{Discussion}

The obtained values of the branching fractions, 
$\br(\rho^0\to\pipig)=\rhobr$
and $\br(\omega\to\pipig)=\ombr$, are in good agreement with the  
previous measurements by GAMS~\cite{gams} and SND~\cite{snd2}. Both 
values are significantly higher than
those predicted by the vector dominance model with the
standard value of the coupling constant: 
$\sim 1 \times 10^{-5}$ and $\sim 3 \times 10^{-5}$~\cite{bgp92}, 
respectively.
An attempt to explain the obtained branching ratios results in a
high value of $g_{\rho\omega\pi}$ contradicting 
the other observations like, e.g., 
the experimental values of the $\omega \to \pi^+\pi^-\pi^0$ 
and $\omega \to \pi^0\gamma$ widths.

Theoretical papers on the subject
\cite{Sin62,Ren69,fo90,bgp92,Pra94,mhot99,belmn01,gs01,gsy01,pho02,gsy03,gky03,ok03}
offer a broad choice of effects influencing the discussed decays.
In addition to the $\omega\rho\pi$ 
%($\rho\omega\pi$) 
transition they include: $\rho-\omega$ mixing, pion and kaon loops,
various scalar ($\fzero, f_0(980), a_0(980)$) and tensor ($f_2(1270)$ and 
$a_2(1320)$) intermediate mesons decaying into $\pi\pi$ $(\eta\pi)$.

Predictions of these models differ rather strongly from each other
reflecting various approaches applied by their authors~\cite{pho02}.
While most of the recent papers agree that the observed value
of the branching fraction for the $\rho^0\to\pipig$ decay can be ascribed
to the intermediate $\fzero$ state, the situation with the corresponding
$\omega$ decay remains controversial.
The corresponding ranges of the predicted values of branching fractions are
summarized in Table~\ref{tabpred}. Note that from the upper limits for the
non-$\omega\pi$ $\pipig$ cross section obtained by us at higher energy in
Ref.~\cite{ompi_cmd} a significant contribution from the $f_0(980)\gamma$ or  
$f_2(1270)\gamma$ mechanisms appears not very likely.

\begin{table*}
\caption{Predictions for branching fractions of $\rho$, 
$\omega\to\pipig$, $\etapig$ decays.}
\medskip
 \begin{tabular*}{\textwidth}{l@{\extracolsep{\fill}}c}
\hline\hline
Mode & Branching fraction \\\hline
$\rho^0 \to \pi^0\pi^0\gamma$ & (1.1--4.7) $\times 10^{-5}$ \\
$\omega \to \pi^0\pi^0\gamma$ & (1.4--8.2) $\times 10^{-5}$ \\
$\rho^0 \to \eta\pi^0\gamma$ & 2$\times 10^{-10}$--4$\times 10^{-6}$ \\
$\omega \to \eta\pi^0\gamma$& 8.3$\times 10^{-8}$--6.3$\times 10^{-6}$ \\
\hline\hline
\end{tabular*} 
\label{tabpred}

\end{table*}

Much higher data samples of the $\rho$ and $\omega$ decays expected
in experiments at the upgraded collider VEPP-2000 in 
Novosibirsk~\cite{vep2000} will help
to significantly improve our understanding of their radiative decays. 

From the obtained results on the cross section of the radiative
processes $\ee\to\pipig$, $\etapig$ one can estimate
a possible contribution of the previously unstudied radiative
processes to the leading order hadronic correction to the muon
anomalous magnetic moment. To this end we first calculate the
contribution of the process  $\ee\to \pi^0\pi^0\gamma$ using the
``bare'' cross section, $\hat{\sigma}$, from Table~\ref{ompi_cs} in the
energy range below 920 MeV. The result contains a piece coming from
the $\omega \to \pi^0\pi^0\gamma$ decay already taken into account
in Ref.~\cite{dav} in the whole $\omega$ meson contribution,
$a^{\omega}_{\mu}=(37.96 \pm 1.07) \cdot 10^{-10}$.
This $\omega$ meson contribution is subtracted from the value above 
using the branching ratio $\br(\omega \to\pipig)$ with the result
$(6.08 \pm 0.82) \cdot 10^{-12}$. 
A possible
contribution from the process $\ee \to \pi^+\pi^-\gamma$ is twice
that of $\ee\to \pipig$, so that    
$$
 {a}^{LO,\pi\pi\gamma}_{\mu}(600~ \rm MeV - 920~ \rm MeV) =(18.2 \pm 2.5) 
\cdot 10^{-12}.
$$ 
Adding the contribution from the $\eta\pi\gamma$ final state, we
finally obtain:
$$
 {a}^{LO,rad}_{\mu}(600~ \rm MeV - 920~ \rm MeV) < 0.24 \cdot 10^{-10}~~
{\rm at}~~90\%\ {\rm CL}.
$$
Adding the upper limit from the energy range 920 MeV -- 2000 MeV
obtained previously~\cite{ompi_cmd}, we obtain
$$
 {a}^{LO,rad}_{\mu}(600~ \rm MeV - 2000~ \rm MeV) < 0.7 \cdot 10^{-10}~~
{\rm at}~~90\%\ {\rm CL}
$$ 
or about 10\% of the current 
uncertainty of $a^{LO,had}_{\mu}$~\cite{dav}.

\section{Conclusions}

The following results are obtained in this work:
\begin{itemize}
\item
Using a data sample corresponding to integrated luminosity of 
7.7~pb$^{-1}$, the cross section of the process $\eepipig$ 
has been measured in the c.m. energy range 600--970~MeV. The
values of the cross section are consistent with those obtained by
the SND detector~\cite{snd2} and have the similar accuracy.
The following branching ratios have been determined:
$\br(\rho\to\pipig)=\rhobr$ and
$\br(\omega\to\pipig)=\ombr$.
\item We confirm evidence for the $\rho\to\fzero\gamma$ decay 
  with the branching fraction
$\br(\rho \to \fzero\gamma)=\rhosibr$  reported
  by the SND Collaboration~\cite{snd2}.
\item
A first search for the process $\ee \to \etapig$ was performed allowing to
set the 90\% CL upper limits: $\sigma(\ee \to \etapig) < 57$~pb in the
c.m. energy range 685--920~MeV and
$\br(\omega\to\etapig) < 3.3 \times 10^{-5}$.
\item
A possible contribution of the studied  radiative processes to the
muon anomalous magnetic moment was estimated to be negligible.  
\end{itemize}

The authors are grateful to the staff of VEPP-2M for the
excellent performance of the collider, and to all engineers and 
technicians who participated in the design, commissioning and operation
of CMD-2. This work is supported in part 
by grants DOE DEFG0291ER40646, NSF PHY-9722600, NSF PHY-0100468,
INTAS 96-0624, RFBR-98-02-1117851 and RFBR-03-02-16843.


\begin{thebibliography}{99}

\bibitem{Sin62} P.~Singer, Phys. Rev. 128 (1962) 2789;\\
Erratum - ibid, 161 (1967) 1694.

\bibitem{Ren69} F.M.~Renard, Nuovo Cimento, 62A (1969) 475.

\bibitem{fo90} S.~Fajfer and R.J.~Oakes, Phys. Rev. D 42 (1990) 2392.

\bibitem{bgp92} A.~Bramon, A.~Grau and G.~Pancheri,
Phys. Lett. B 283 (1992) 416; \\
 A.~Bramon, A.~Grau and G.~Pancheri,
Phys. Lett. B 289 (1992) 97. 

\bibitem{Pra94} J.~Prades, Z. Phys. C 63 (1994) 491.

\bibitem{mhot99} E.~Marco, S.~Hirenzaki, E.~Oset and H.~Toki,
Phys. Lett. B 470 (1999) 20. 

\bibitem{belmn01} A.~Bramon, R.~Escribano, J.L.~Lucio Martinez and 
M.~Napsuciale,
Phys. Lett. B 517 (2001) 345.

\bibitem{gs01} D.~Guetta and P.~Singer,
Phys. Rev. D 63 (2001) 017502.

\bibitem{gsy01} A.~Gokalp, Y.~Sarac and O.~Yilmaz,
Eur. Phys. J. C 22 (2001) 327.

\bibitem{pho02} J.E.~Palomar, S.~Hirenzaki and E.~Oset,
Nucl. Phys. A 707 (2002) 161. 

\bibitem{gsy03} A.~Gokalp, S.~Solmaz and O.~Yilmaz,
Phys. Rev. D 67 (2003) 073007.

\bibitem{gky03} A.~Gokalp, A.~Kucukarslan and O.~Yilmaz,
Phys. Rev. D 67 (2003) 073008.

\bibitem{ok03} Y.~Oh and H.~Kim, hep-ph/007286, July 2003.

\bibitem{sndphi}{M.N.~Achasov et al.,
Phys. Lett. B 438 (1998) 441. \\
M.N.~Achasov et al., JETP Lett. 68 (1998) 573. \\
M.N.~Achasov et al., Phys. Lett. B 440 (1998) 442.}

\bibitem{cmd2phi} {R.R.~Akhmetshin et al., 
Phys. Lett. B 462 (1999) 380.}

\bibitem{kloe} {A.~Aloisio et al., Phys. Lett. B 536 (2002) 209. \\
A.~Aloisio et al., Phys. Lett. B 537 (2002) 21.}

\bibitem{pdg}{
K.~Hagiwara et al., Phys. Rev. D 66 (2002) 010001.}

\bibitem{ndrho}{I.B.~Vasserman et al., Sov. J. Nucl. Phys. 47 (1988)
1035; \\
S.I.~Dolinsky et al., Phys. Rep. 202 (1991) 99.  }

\bibitem{gams} {D.M.~Alde et al., Phys. Lett. B 340 (1994) 122.}

\bibitem{snd1}{M.N.~Achasov et al.,
JETP Lett. 71 (2000) 355.}

\bibitem{snd2}{M.N.~Achasov et al.,
Phys. Lett. B 537 (2002) 201.}

\bibitem{ompi_cmd} {R.R.~Akhmetshin et al., 
Phys. Lett. B 562 (2003) 173.}

\bibitem{cmddet} {G.~A.~Aksenov et al., 
Preprint Budker INP 85-118, Novosibirsk, 1985; \\
E.~V.~Anashkin et al., ICFA Instr. Bulletin 5 (1988) 18.}

\bibitem{prep} {R.~R.~Akhmetshin et al., 
Preprint Budker INP 99-11, Novosibirsk, 1999.}

%\bibitem{mast}
%{P.~P.~Krokovny, Master's Thesis, Novosibirsk State University, 2000.}
 
\bibitem{feldman}{G.J.~Feldman and R.D.~Cousins,
 Phys. Rev. D  57 (1998) 3873.}
 
\bibitem{rho} {R.R.~Akhmetshin et al., 
Phys. Lett. B 527 (2002) 161.}

\bibitem{radcor}{E.~A.~Kuraev and V.S.~Fadin, 
Sov. J. Nucl. Phys., 41 (1985) 466.}

%\bibitem{gounaris}{G.~J.~Gounaris and J.J.~Sakurai, 
%Phys. Rev. Lett. 21 (1968) 244.} 

\bibitem{vep2000}{ Yu.M.~Shatunov, Proc. of EPAC 2000, Vienna,
Austria, 2000, p.439.}

%\bibitem{rt03} {M.P.~Rekalo and E.~Tomasi-Gustafsson, 
%Nucl. Phys. A 725 (2003) 116.}

\bibitem{dav} {M.~Davier, S.~Eidelman, A.~H\"ocker, Z.~Zhang, 
Eur. Phys. J. C 27 (2003) 497;
M.~Davier, S.~Eidelman, A.~H\"ocker, Z.~Zhang, hep-ph/0308213.}
\end{thebibliography}
\end{document}